\documentclass[aps,pre,reprint,twocolumn,superscriptaddress,showpacs,floatfix,longbibliography]{revtex4-1}

\usepackage{graphicx}
\usepackage{amsmath,latexsym}
\usepackage{amssymb}
\usepackage{epsf}
\usepackage{color}

\begin{document}

\title{Information transmission in a two-step cascade: Interplay of activation and repression}

\author{Tuhin Subhra Roy}
\affiliation{Department of Chemistry, Bose Institute, 93/1 A P C Road, Kolkata 700009, India}

\author{Mintu Nandi}
\affiliation{Department of Chemistry, University of Calcutta, 92 A P C Road, Kolkata 700009, India}

\author{Ayan Biswas}
\affiliation{Department of Chemistry, Bose Institute, 93/1 A P C Road, Kolkata 700009, India}

\author{Pinaki Chaudhury}
\email{pcchem@caluniv.ac.in}
\affiliation{Department of Chemistry, University of Calcutta, 92 A P C Road, Kolkata 700009, India}

\author{Suman K Banik}
\email{skbanik@jcbose.ac.in}
\affiliation{Department of Chemistry, Bose Institute, 93/1 A P C Road, Kolkata 700009, India}

\date{\today}

\begin{abstract}
We present an information-theoretic formalism to study signal transduction in four architectural variants of a model two-step cascade with increasing input population. Our results categorize these four types into two classes depending upon the effect played out by activation and repression on mutual information, net synergy, and signal-to-noise ratio. Within the Gaussian framework and using the linear noise approximation, we derive the analytic expressions for these metrics to establish their underlying relationships in terms of the biochemical parameters. We also verify our approximations through stochastic simulations.
\end{abstract}

\maketitle


\section{Introduction}

Signal transduction is an essential process for a living system to survive in a dynamic environment. By the mechanism of signal transduction, a living entity senses the changes in its immediate surroundings. While doing so, a single or a multi-cellular species utilizes biochemical pathways to process the environmental information \cite{Alon2006,Alon2007}. Information processing within the purview of the gene regulatory network (GRN) has attracted lots of attention recently. Starting from simple motif to complex networks, researchers have addressed key issues related to information processing capacity, fidelity, and functionality of the network of interest \cite{Cheong2011,Bowsher2014,Mahon2014,Selimkhanov2014,Hansen2015,Jost2020}.

Quantification of information processing can be achieved using the tools of information theory formulated by Shannon \cite{Shannon1948,Shannon1963}. Shannon's formalism provides the most used measure, mutual information (MI) $I (i; j)$ defined between two random variables $i$ and $j$ \cite{Cover2006}
\begin{eqnarray}
\label{eq00}
I (i; j) = \sum_{i,j} p(i, j) \log_2 \left [ \frac{p(i, j)}{p(i) p(j)} \right ]. 
\end{eqnarray}
Here, $p(i)$ and $p(j)$ are the marginal and $p(i,j)$ is the joint probability function. MI is measured in bits due to base 2 in the logarithmic function. The above relation suggests that a probabilistic interpretation is necessary for variables of interest. This essentially leads to the study of cellular dynamics in a fluctuating environment. To calculate MI, one thus needs a stochastic description of the signal transduction mechanism that mimics the dynamics in a single cell environment.

In general, MI is a measure of the correlation between two random variables whose entropy or fluctuation spaces overlap. From Eq.~(\ref{eq00}), it can be easily inferred that $I(i;j) \geqslant 0$, the lower bound being realized whenever $p(i,j)=p(i)p(j)$ or in other words, correlation is lost between $i~\text{and}~j$, i.e., these random variables become independent of each other. MI is a special case of what is generally known to mathematicians; relative entropy or the Kullback-Leibler (KL) distance which quantifies the \textit{distance} between two probability spaces but does not meet the requirement set by the triangle inequality. But unlike the KL distance, MI by construction is symmetric with respect to its argument variables i.e., $I(i;j)=I(j;i)$ \cite{Cover2006}. Due to this feature, MI on its own can not provide insights into the temporal structure of processed information by a set of random variables. From the entropy perspective, the following relations hold between MI and corresponding marginal ($H(i), H(j)$), joint ($H(i,j)$), and the conditional ($H(i|j), H(j|i)$) entropy \cite{Cover2006}. 
\begin{subequations}
\begin{eqnarray}
I(i;j) & = & H(i)-H(i|j), \label{eq01} \\
& = & H(j)-H(j|i), \label{eq02} \\
& = & H(i)+H(j)-H(i,j). \label{eq03}
\end{eqnarray}

\noindent Where, the entropy definitions are
\begin{eqnarray}
H(i) & = & -\sum_{i} p(i) \log_2 p(i), \label{eq04} \\
H(i,j) & = & -\sum_{i,j} p(i,j) \log_2 p(i,j), \label{eq05} \\
H(i|j) & = & -\sum_{i,j} p(i,j) \log_2 p(i|j). \label{eq06}
\end{eqnarray}
\end{subequations}

\noindent Here, $p(i|j)$ is the conditional probability function. Eqs.~(\ref{eq01}-\ref{eq03}), in essence, demonstrate the fact that MI quantifies the average reduction in entropy of one variable when knowledge of the other one is given.


\begin{figure}[!t]
\includegraphics[width=1.0\columnwidth,angle=0]{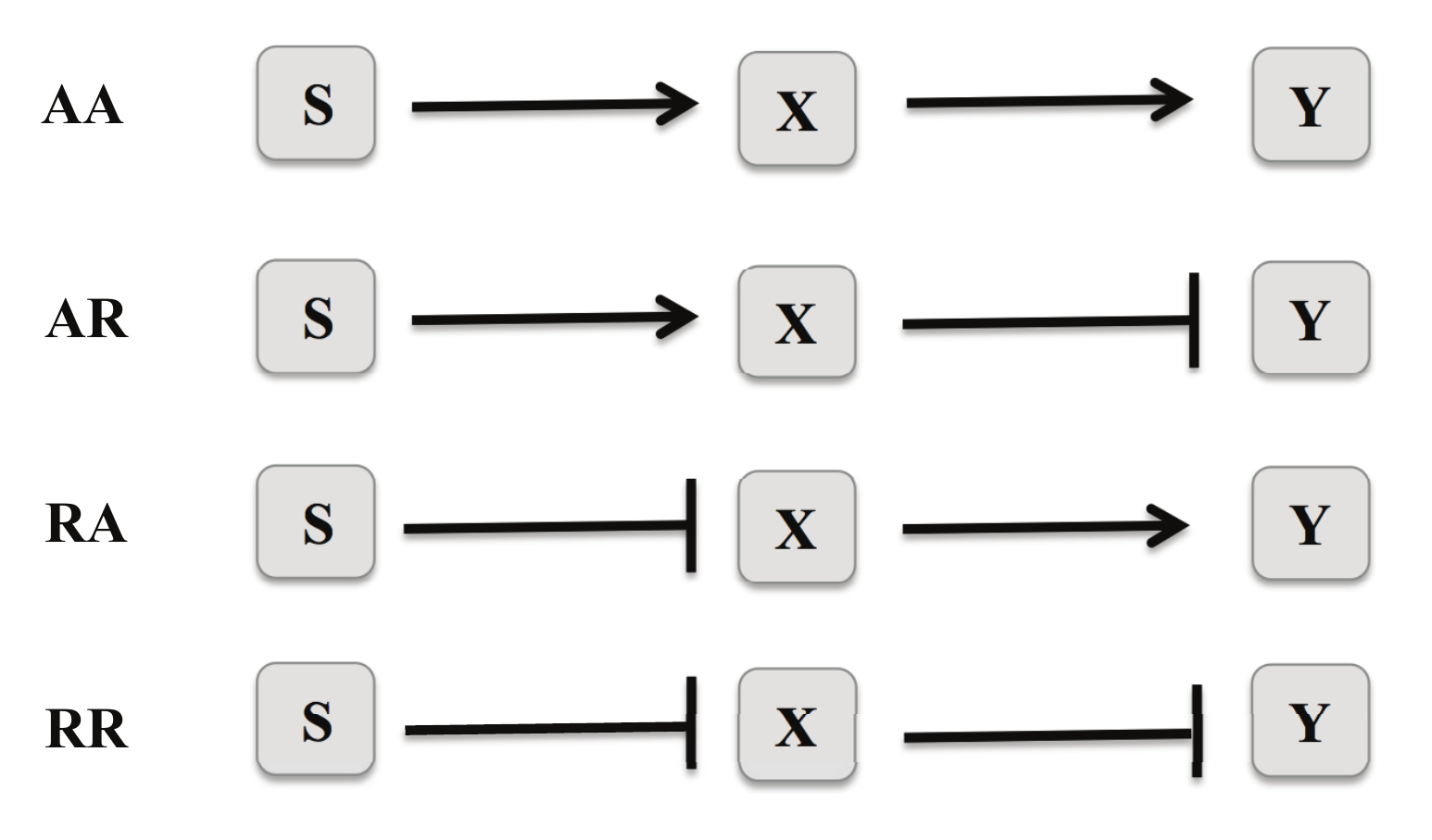}
\caption{Schematic diagram of four different TSC motifs, activation-activation (AA), activation-repression (AR), repression-activation (RA) and repression-repression (RR). In the diagram, $\rightarrow$ and $\dashv$ stand for activation and repression, respectively.
}
\label{fig1}
\end{figure}

The extension of MI into the multivariate realm is straight-forward and is utilized in the present paper. Categorizing the associated random variables into sources and targets of information takes us to ask whether assembling more than one source variable simultaneously can benefit us in extracting more information than what is possible otherwise just by adding their contributions. For a three-variable case, the difference between the three-variable total MI and sum of the two constituent two-variable MIs is often used to quantify the information independence between the source variables \cite{Schneidman2003}. Since this metric characterizes the effective synergistic aspect between the information sources and can be shown to be equal to the difference between synergistic and redundant information according to the formalism of partial information decomposition (PID) \cite{Williams2010}, it is often dubbed as the net synergy \cite{Barrett2015}.

For a system comprised of random variables $i, j$ and $k$ the three variable mutual information $I(i; j,k)$ is written as
\begin{equation}
\label{eq07}
I(i; j,k) = \sum_{i} \sum_{j,k} p(i,j,k) \log_2 
\left [ \frac{p(i,j,k)}{p(i) p(j,k)} \right ],
\end{equation}

\noindent whose entropy representation is given by \cite{Cover2006,Maity2017}
\begin{equation}
\label{eq08}
I(i; j,k) = H(i) + H(j,k) - H(i,j,k).
\end{equation}

\noindent Now considering $i$ as a target variable and $j$ and $k$ as source variables one can rewrite Eq.~(\ref{eq07}) as \cite{Barrett2015}
\begin{subequations}
\begin{equation}
\label{eq09}
I(i; j,k) =  U(i; j|k) + U(i; k|j) + S(i; j,k) + R(i; j,k),
\end{equation}

\noindent where $U(i; j|k)$ is the unique information between the source variable $j$ and the target variable $i$ given the other source variable is $k$. On a similar note $U(i; k|j)$ is the unique information between the source variable $k$ and the target variable $i$, out of two source variables $j$ and $k$. The synergistic information $S(i; j,k)$ is the information of the target variable $i$ when the knowledge of both source variables $j$ and $k$ are known. When knowledge of either of the source variables $j$ and $k$ is sufficient to predict the nature of the target variable $i$, one can quantify the redundant information $R(i; j,k)$. Similarly, two variable mutual information $I(i;j)$ and $I(i;k)$can be written in terms of unique and redundant information
\begin{eqnarray}
\label{eq010}
I(i;j) & = & U(i; j|k) + R(i; j,k), \\  
\label{eq011}
I(i;k) & = & U(i; k|j) + R(i; j,k).
\end{eqnarray}

Following the formalism of PID, one can define the measure net synergy $\Delta I(i; j,k)$ using the above expressions of three variable and two variable mutual information \cite{Williams2010,Barrett2015}
\begin{eqnarray}
\label{eq012}
\Delta I(i; j,k) & =: & I(i; j,k) - I(i;j) - I(i;k), \\
\label{eq013}
& = & S(i; j,k) - R(i; j,k).
\end{eqnarray}
\end{subequations}

\noindent The above definition of net synergy suggests that for a multivariate system the bivariate or the trivariate (multivariate) mutual information alone does not provide useful information about the system. However, an algebraic combination of these measures gives insightful information regarding the process. Depending on the nature of interaction among the nodes of a network $\Delta I$ can take three different values, i.e., $>0$, $<0$ and $=0$ \cite{Schneidman2003}. For $\Delta I > 0$, the synergistic effect overpowers the redundant effect. In other words, common knowledge of both source variables is more effective than the knowledge of a single source variable. Similarly, $\Delta I < 0$ focuses on redundant information sharing between target and source variables. The scenario $\Delta I = 0$ leads to the borderline situation where both common information sharing balances redundant information sharing between source and target variables leading to information independence. We note that classification of net synergy $\Delta I$ has been developed in the context of population coding in neuroscience \cite{Schneidman2003}, however, its application in GTRN is limited.


\begin{table}
\caption{List of functions used in four different TSC motifs.}
\label{table1}
\begin{ruledtabular}
\begin{tabular}{cccc}
Motif & $f_s(s)$ & $f_x(s,x)$ & $f_y(s,x,y)$ \\
\hline
AA & $k_s$ & $k_x \frac{s^n}{K_1^n + s^n}$ & $k_y \frac{x^n}{K_2^n + x^n}$ \\
(${\rm S} \rightarrow {\rm X} \rightarrow {\rm Y}$) & & \\
\\
AR & $k_s$ & $k_x \frac{s^n}{K_1^n + s^n}$ & $k_y \frac{K_2^n}{K_2^n + x^n}$ \\
(${\rm S} \rightarrow {\rm X} \dashv {\rm Y}$) & & \\
\\
RA & $k_s$ & $k_x \frac{K_1^n}{K_1^n + s^n}$ & $k_y \frac{x^n}{K_2^n + x^n}$ \\
(${\rm S} \dashv {\rm X} \rightarrow {\rm Y}$) & & \\
\\
RR & $k_s$ & $k_x \frac{K_1^n}{K_1^n + s^n}$ & $k_y \frac{K_2^n}{K_2^n + x^n}$ \\
(${\rm S} \dashv {\rm X} \dashv {\rm Y}$) & & \\
\end{tabular}
\end{ruledtabular}
\end{table}

The present paper focuses on this principle information-theoretic metric in the perspective of model GRNs, and the next section shows how this can be computed analytically in terms of variances and covariances using a well-known approximation technique. The bacterial GRNs are often found to be enriched with small modular structures named as network motifs \cite{Alon2006}. A two-step cascade (TSC) may be regarded as one prime example of this category. In a TSC, one transcription factor (TF) regulates another TF which finally regulates the output gene-expression, which may result in the production of key enzymes and proteins with a possibility for the latter species acting as regulators in other motifs. Regarding genes/gene-products as nodes and transcriptional regulations (activation/repression) as edges have simplified the gene-regulation phenomenon into a network analysis problem. To this end we have considered four types of TSC depending on the possible interaction topology between the nodes, namely, activation-activation (AA), activation-repression (AR), repression-activation (RA) and repression-repression (RR) (see Fig.~\ref{fig1}).

In one of our earlier reports \cite{Biswas2016}, we have investigated the responses of net synergy and the signal-to-noise ratio (SNR) as a function of input degradation rate parameter in an AA type of TSC with fixed population levels. In the current manuscript, we intend to extend the analysis to showcase those metrics' responses in all of the possible topologies of the generic TSC network with tunable signaling strength. In the following, we have studied information processing in four different architectures available to a generic TSC while paying attention to crucial metrics like MI, net synergy, and SNR. Our observations demonstrate the categoric influence played out by transcriptional regulation on fluctuation propagation in TSC pathways. The present work also aims to highlight the dependence of signal fidelity on the interaction topology of a TSC. Moreover, the present construct verifies the role of the net synergy as an indicator of the signaling efficiency in midst of the stochasticity due to the interaction topologies of the TSC.


\section{Model and Methods}

The Langevin equations governing the dynamics of a generic TSC are \cite{Biswas2016},
\begin{eqnarray}
\label{eq1}
\frac{ds}{dt} & = & f_{s}(s)-\mu_s s+\xi_s(t), \\
\label{eq2}
\frac{dx}{dt} & = & f_{x}(s,x)-\mu_x x+\xi_x(t), \\
\label{eq3}
\frac{dy}{dt} & = & f_{y}(s,x,y)-\mu_y y+\xi_y(t).
\end{eqnarray}

\noindent In Eqs.~(\ref{eq1}-\ref{eq3}), $s, x$ and $y$ are the copy numbers of species S, X and Y, respectively, expressed in molecules/$V$. Here, $V$ is the unit effective cellular volume. Here copy number stands for population of a species. $f_{s}(s)$, $f_{x}(s, x)$ and $f_{y}(s, x, y)$ are the production term associated with $s$, $x$ and $y$, respectively. We note that the production term of each component, in general, are nonlinear in nature \cite{Bintu2005,Ziv2007,Tkacik2008a,Tkacik2008b,Ronde2012}. The degradation rate parameters associated with $s$, $x$ and $y$ are written as $\mu_s$, $\mu_x$, and $\mu_y$, respectively. The constants $\mu_s$, $\mu_x$, and $\mu_y$ correspond to the inverse lifetime of the gene product of the respective component.

The noise processes $\xi_i (t)$ considered here are independent and Gaussian distributed with properties $\langle \xi_{i}(t) \rangle$ = 0 and $\langle \xi_{i}(t)\xi_{j}(t^{'}) \rangle$ = $\langle |\xi_{i}|^{2}\rangle \delta_{ij} \delta(t-t^{'})$ with $\langle |\xi_{i}|^{2}\rangle = \langle f_{i} \rangle + \mu_{i}\langle i \rangle = 2\mu_{i}\langle i \rangle$ ($i =s, x$, and $y$) at steady state \cite{Elf2003,Swain2004,Paulsson2004,Tanase2006,Warren2006,Kampen2007,Mehta2008,Ronde2010}. The nature of noise correlation also suggests that the noise processes are uncorrelated with each other \cite{Swain2016}. Here, $\langle \cdots \rangle$ represents steady state ensemble average. The noise processes we consider here are intrinsic in nature. As a result, variability induced in the dynamics are controlled by the intrinsic noise processes although it has been observed that fluctuations in gene expression in a genetically identical population are mostly controlled by extrinsic noise processes \cite{Elowitz2002,Raser2004}. Depending on the nature of interactions, activation or repression shown in Fig.~\ref{fig1}, we consider specific form of the functions $f_s (s)$, $f_x (s, x)$ and $f_y (s, x, y)$ (see Table~\ref{table1}).

We now employ linear noise approximation (LNA) \cite{Elf2003,Kampen2007} to calculate the second moments associated with $s, x$ and, $y$. To this end we use perturbation of linear order $\delta u(t) = u(t) - \langle u \rangle$ where $\langle u \rangle$ is the average population of $u$ at steady state and recast Eqs.~(\ref{eq1}-\ref{eq3}) in the following form
\begin{equation}
\label{eq4}
\frac{d\mathbf{\delta W}}{dt} = \mathbf{J}_{W=\langle W \rangle}
\mathbf{\delta W}(t) + \mathbf{\Xi}(t).
\end{equation}

\noindent In the above equation, $\mathbf{\delta W}(t)$ denotes the fluctuation matrix containing the linear order perturbation terms. The noise matrix $\mathbf{ \Xi}(t)$ along with $\mathbf{\delta W}(t)$ can be written as
\begin{eqnarray*}
\mathbf{\delta W}(t) = \left( 
\begin{array}{ccc} \delta s(t) \\ \delta x(t) \\ \delta y(t) \\
\end{array} \right),
\mathbf{ \Xi}(t) = \left( 
\begin{array}{ccc}  \xi_{s}(t) \\  \xi_{x}(t) \\  \xi_{y}(t) \\
\end{array} \right).
\end{eqnarray*}

\noindent
$\mathbf{J}$ represents the Jacobian matrix at steady state,
\begin{eqnarray*}
\mathbf{J} = \left [
\begin{array}{ccc}
f^{\prime}_{s,s} (\langle s \rangle) - \mu_s & 0 & 0 \\
f^{\prime}_{x,s} (\langle s \rangle, \langle x \rangle) &
f^{\prime}_{x,x} (\langle s \rangle, \langle x \rangle) - \mu_x & 0 \\
f^{\prime}_{y,s} (\langle s \rangle, \langle x \rangle, \langle y \rangle) &
f^{\prime}_{y,x} (\langle s \rangle, \langle x \rangle, \langle y \rangle) &
f^{\prime}_{y,y} (\langle s \rangle, \langle x \rangle, \langle y \rangle) - \mu_y
\end{array}
\right ].
\end{eqnarray*}

\noindent Here $f^{\prime}_{s,s}$ stands for differentiation of $f_s$ with respect to $s$ and evaluated at $s = \langle s \rangle$, and so on. The variance and covariance is evaluated using the Lyapunov equation at steady state \cite{Keizer1987,Elf2003,Paulsson2004,Paulsson2005,Kampen2007}
\begin{equation}
\label{eq5}
\mathbf{J \Sigma} + \mathbf{\Sigma J}^{T} + \mathbf{D} = \mathbf{0}.
\end{equation}

\noindent In Lyapunov equation, $\mathbf{\Sigma}$ and $\mathbf{D}$ stand for covariance matrix and diffusion matrix, respectively. The diffusion matrix $\mathbf{D}$ incorporates different noise strength using the relation $\mathbf{D} = \langle \mathbf{\Xi} \mathbf{\Xi}^T \rangle$. As the noise processes $\xi_s$, $\xi_x$ and $\xi_y$ are uncorrelated, the off-diagonal elements of the diffusion matrix $\mathbf{D}$ are zero. Here $T$ represents transpose of a matrix. Analytical solution of Eq.~(\ref{eq5}) yields variance and covariance associated with different TSC motifs (see Table~\ref{table2}).

The variance and covariance associated with different TSC motifs given in Table~\ref{table2} can be used to calculate the two and three variable MIs using Gaussian framework \cite{Shannon1948,Cover2006,Barrett2015},
\begin{eqnarray}
\label{eq014}
I(s; x) & = & \frac{1}{2} \log_2 \left [ 
\frac{\Sigma (s) \Sigma(x)}{\Sigma (s) \Sigma(x) - \Sigma^2(s,x)}
\right ], \\
\label{eq015}
I(s; y) & = & \frac{1}{2} \log_2 \left [ 
\frac{\Sigma (s) \Sigma(y)}{\Sigma (s) \Sigma(y) - \Sigma^2(s,y)}
\right ], \\
\label{eq016}
I(s; x,y) & = & \frac{1}{2} \log_2 \left [ 
\frac{\Sigma (s) (\Sigma(x) \Sigma(y) - \Sigma^2(x,y))}{| \Delta |}
\right ],
\end{eqnarray}

\noindent
where
\begin{eqnarray*}
\Delta = \left (
\begin{array}{ccc}
\Sigma(s) & \Sigma(s,x) & \Sigma(s,y) \\
\Sigma(s,x) & \Sigma(x) & \Sigma(x,y) \\
\Sigma(s,y) & \Sigma(x,y) & \Sigma(y)
\end{array}
\right ).
\end{eqnarray*}

\noindent In the above expressions, $\Sigma(s)$, $\Sigma(x)$ and $\Sigma(y)$ are the variance of $s$, $x$ and $y$, respectively. The covariance between $s$ and $x$, $s$ and $y$, and $x$ and $y$ are written as $\Sigma(s, x)$, $\Sigma(s, y)$ and $\Sigma(x, y)$, respectively.


\begin{widetext}

\begingroup
\squeezetable
\begin{table}
\caption{Variance and covariance of $s$, $x$ and $y$ in four different TSC motifs. $\Sigma(s)$, $\Sigma(x)$ and $\Sigma(y)$ are the variance of $s$, $x$ and $y$, respectively. The covariance between $s$ and $x$, $s$ and $y$, and $x$ and $y$ are written as $\Sigma(s, x)$, $\Sigma(s, y)$ and $\Sigma(x, y)$, respectively.
}
\label{table2}
\begin{ruledtabular}
\begin{tabular}{ccccccc}
Motif & $\Sigma (s)$ & $\Sigma (s,x)$ & $\Sigma (s,y)$ & $\Sigma (x)$ & $\Sigma (x,y)$ & $\Sigma (y)$ \\
\hline
AA & $\langle s \rangle$ & 
$\frac{n k_x K_1^n \langle s \rangle^n}{(K_1^n + \langle s \rangle^n)^2 (\mu_s + \mu_x)}$ &
$\frac{n k_y K_2^n \langle x \rangle^{n-1} \Sigma (s,x)}{(K_2^n + \langle x \rangle^n)^2 (\mu_s + \mu_y)}$ &
$\langle x \rangle + \frac{n k_x K_1^n \langle s \rangle^{n-1} \Sigma (s,x)}{(K_1^n + \langle s \rangle^n)^2 \mu_x }$ &
$\frac{n k_y K_2^n \langle x \rangle^{n-1} \Sigma (x)}{(K_2^n + \langle x \rangle^n)^2 (\mu_x + \mu_y)}$ &
$\langle y \rangle + \frac{n k_y K_2^n \langle x \rangle^{n-1} \Sigma (x,y)}{(K_2^n + \langle x \rangle^n)^2 \mu_y }$ \\
& & & & & +$\frac{n k_x K_1^n \langle s \rangle^{n-1} \Sigma (s,y)}{(K_1^n + \langle s \rangle^n)^2 (\mu_x + \mu_y)}$ & \\
\\
AR & $\langle s \rangle$ & 
$\frac{n k_x K_1^n \langle s \rangle^n}{(K_1^n + \langle s \rangle^n)^2 (\mu_s + \mu_x)}$ &
-$\frac{n k_y K_2^n \langle x \rangle^{n-1} \Sigma (s,x)}{(K_2^n + \langle x \rangle^n)^2 (\mu_s + \mu_y)}$ &
$\langle x \rangle + \frac{n k_x K_1^n \langle s \rangle^{n-1} \Sigma (s,x)}{(K_1^n + \langle s \rangle^n)^2 \mu_x }$ &
-$\frac{n k_y K_2^n \langle x \rangle^{n-1} \Sigma (x)}{(K_2^n + \langle x \rangle^n)^2 (\mu_x + \mu_y)}$ &
$\langle y \rangle - \frac{n k_y K_2^n \langle x \rangle^{n-1} \Sigma (x,y)}{(K_2^n + \langle x \rangle^n)^2 \mu_y }$ \\
& & & & & +$\frac{n k_x K_1^n \langle s \rangle^{n-1} \Sigma (s,y)}{(K_1^n + \langle s \rangle^n)^2 (\mu_x + \mu_y)}$ & \\
\\
RA & $\langle s \rangle$ & 
-$\frac{n k_x K_1^n \langle s \rangle^n}{(K_1^n + \langle s \rangle^n)^2 (\mu_s + \mu_x)}$ &
$\frac{n k_y K_2^n \langle x \rangle^{n-1} \Sigma (s,x)}{(K_2^n + \langle x \rangle^n)^2 (\mu_s + \mu_y)}$ &
$\langle x \rangle - \frac{n k_x K_1^n \langle s \rangle^{n-1} \Sigma (s,x)}{(K_1^n + \langle s \rangle^n)^2 \mu_x }$ &
$\frac{n k_y K_2^n \langle x \rangle^{n-1} \Sigma (x)}{(K_2^n + \langle x \rangle^n)^2 (\mu_x + \mu_y)}$ &
$\langle y \rangle + \frac{n k_y K_2^n \langle x \rangle^{n-1} \Sigma (x,y)}{(K_2^n + \langle x \rangle^n)^2 \mu_y }$ \\
& & & & & -$\frac{n k_x K_1^n \langle s \rangle^{n-1} \Sigma (s,y)}{(K_1^n + \langle s \rangle^n)^2 (\mu_x + \mu_y)}$ & \\
\\
RR & $\langle s \rangle$ & 
-$\frac{n k_x K_1^n \langle s \rangle^n}{(K_1^n + \langle s \rangle^n)^2 (\mu_s + \mu_x)}$ &
-$\frac{n k_y K_2^n \langle x \rangle^{n-1} \Sigma (s,x)}{(K_2^n + \langle x \rangle^n)^2 (\mu_s + \mu_y)}$ &
$\langle x \rangle - \frac{n k_x K_1^n \langle s \rangle^{n-1} \Sigma (s,x)}{(K_1^n + \langle s \rangle^n)^2 \mu_x }$ &
-$\frac{n k_y K_2^n \langle x \rangle^{n-1} \Sigma (x)}{(K_2^n + \langle x \rangle^n)^2 (\mu_x + \mu_y)}$ &
$\langle y \rangle - \frac{n k_y K_2^n \langle x \rangle^{n-1} \Sigma (x,y)}{(K_2^n + \langle x \rangle^n)^2 \mu_y }$ \\
& & & & & -$\frac{n k_x K_1^n \langle s \rangle^{n-1} \Sigma (s,y)}{(K_1^n + \langle s \rangle^n)^2 (\mu_x + \mu_y)}$ & \\
\end{tabular}
\end{ruledtabular}
\end{table}
\endgroup

\end{widetext}

We note here that calculations presented up to Eq.~(\ref{eq5}) are applicable for Hill coefficient $n \geqslant 1$ within the purview of LNA. In the next section, we have presented results for $n = 1$ and $n = 2$. For $n > 1$, say $n = 2$, etc., will bring in higher levels of nonlinearity in the dynamics. We have, however, excluded the phenomenon of gene switching, autoregulation (positive/negative), and cross-regulation to avoid bimodal distribution of system variables. Our model also does not show the generation of noise-induced bistability for an otherwise deterministic monostable system \cite{To2010}.

The kinetics of signal transmission in all the TSC motifs are simulated using stochastic simulation algorithm \cite{Gillespie1976,Gillespie1977}. The associated kinetic schemes and propensities used for numerical simulation are given in Table~\ref{table3}. Table~\ref{table3} suggests that even for $n = 1$ the propensities used are of Hill type. The propensities used in the present work are not derived from the first principle and hence may lead to misleading results even under the purview of LNA \cite{Thomas2012}. The limitation of using intrinsic noise, Hill coefficient $n = 1$, and phenomenological propensities within the framework of LNA could be overcome by the recent extension of LNA \cite{Thomas2014,Keizer2019}.
In our calculation, we have used $\mu_s < \mu_x < \mu_y$ for maximum information flow due to separation of time scale \cite{Maity2015}. To this end we used $\mu_s=0.1$ sec$^{-1}$, $\mu_x=1$ sec$^{-1}$, $\mu_y=10$ sec$^{-1}$. We set $K_1 = 50$ (molecules/V) and $K_2 = 100$ (molecules/V). The mean copy numbers of $\langle s \rangle$, $\langle x \rangle$ and $\langle y \rangle$ are 50, 100 and 100, respectively, with unit molecules/V. The average population of signal $\langle s \rangle$ is tuned to compute different statistical measures.  Statistical averaging is done using the steady state output of $10^6$ independent trajectories.


\begingroup
\squeezetable
\begin{table}[!b]
\caption{Table of kinetics and associated propensities for different TSC motifs.}
\label{table3}
\begin{tabular}{|ccc|ccc|}
\hline
 & AA & & & AR & \\
\hline
Kinetics & Rate Constant & Propensity & Kinetics & Rate Constant & Propensity \\
\hline
$\phi \rightarrow$ S & $k_s$ & $k_s$ & $\phi \rightarrow$ S & $k_s$ & $k_s$ \\
S $\rightarrow \phi$ & $\mu_s$ & $\mu_s s$ & S $\rightarrow \phi$ & $\mu_s$ & $\mu_s s$ \\
S $\rightarrow$ S+X & $k_x$ & $k_x \frac{s^n}{K_1^n+s^n}$ & S $\rightarrow$ S+X & $k_x$ & $k_x \frac{s^n}{K_1^n+s^n}$ \\
X $\rightarrow \phi$ & $\mu_x$ & $\mu_x x$ & X $\rightarrow \phi$ & $\mu_x$ & $\mu_x x$ \\
X $\rightarrow$ X+Y & $k_y$ & $k_y \frac{x^n}{K_2^n+x^n}$ & X $\rightarrow$ X+Y & $k_y$ & $k_y \frac{K_2^n}{K_2^n+x^n}$ \\
Y $\rightarrow \phi$ & $\mu_y$ & $\mu_y y$ & Y $\rightarrow \phi$ & $\mu_y$ & $\mu_y y$ \\
\hline
 & RA & & & RR & \\
\hline
Kinetics & Rate Constant & Propensity & Kinetics & Rate Constant & Propensity \\
\hline
$\phi \rightarrow$ S & $k_s$ & $k_s$ & $\phi \rightarrow$ S & $k_s$ & $k_s$ \\
S $\rightarrow \phi$ & $\mu_s$ & $\mu_s s$ & S $\rightarrow \phi$ & $\mu_s$ & $\mu_s s$ \\
S $\rightarrow$ S+X & $k_x$ & $k_x \frac{K_1^n}{K_1^n+s^n}$ & S $\rightarrow$ S+X & $k_x$ & $k_x \frac{K_1^n}{K_1^n+s^n}$ \\
X $\rightarrow \phi$ & $\mu_x$ & $\mu_x x$ & X $\rightarrow \phi$ & $\mu_x$ & $\mu_x x$ \\
X $\rightarrow$ X+Y & $k_y$ & $k_y \frac{x^n}{K_2^n+x^n}$ & X $\rightarrow$ X+Y & $k_y$ & $k_y \frac{K_2^n}{K_2^n+x^n}$ \\
Y $\rightarrow \phi$ & $\mu_y$ & $\mu_y y$ & Y $\rightarrow \phi$ & $\mu_y$ & $\mu_y y$ \\
\hline
\end{tabular}
\end{table}
\endgroup

For biochemical kinetics under the influence of Gaussian noise processes, one can make use of different variance and covariance to define two variable and three variable MI. These MIs are then used to calculate net synergy for the TSC motif \cite{Schneidman2003,Barrett2015}
\begin{equation}
\label{eq6}
\Delta I(s; x,y) =: I(s; x,y)-I(s; x)-I(s; y) .
\end{equation}

\noindent Earlier analysis \cite{Biswas2016} of a TSC system reveals that $I(s; x,y) \approx I(s; x)$ and $I(s; x) \geqslant I(s; y)$ due to data processing inequality (DPI) and Markov chain property of the linear architecture \cite{Cover2006}. To be specific for a TSC $I(s; x) > I(s; y)$. Considering this Eq.~(\ref{eq6}) can be written as
\begin{eqnarray}
\label{eq7}
\Delta I(s; x,y) & \approx & - I(s; y) \nonumber \\
& = & 
- \frac{1}{2} \log_2 \left [ 1 + \frac{\Sigma^2 (s,y)}{\Sigma (s) \Sigma (y) - \Sigma^2 (s,y)} \right ],
\end{eqnarray}

\noindent where the different MI terms are expressed in the unit of bits due to the base 2 in the logarithmic function. We note here that the quantity $\Sigma^2 (s,y)/[\Sigma (s) \Sigma (y) - \Sigma^2 (s,y)]$ is defined as signal-to-noise ratio (SNR). SNR provides a measure of the fidelity of the signaling cascade \cite{Cheong2011,Bowsher2013}. A high value of SNR between the input and the output reveals high signal processing capacity of a biochemical network.


\begin{figure}[!t]
\includegraphics[width=1.0\columnwidth,angle=0]{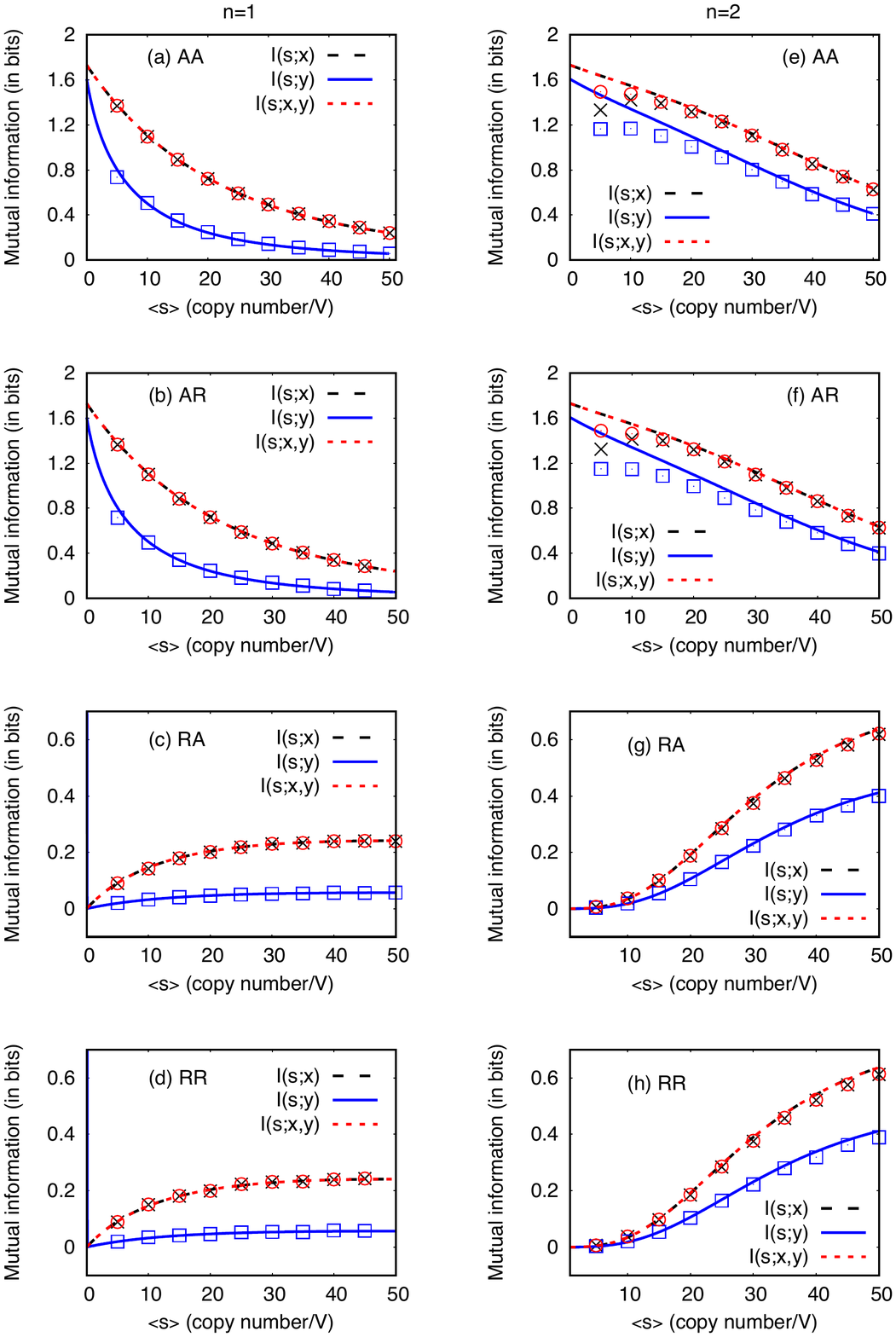}
\caption{(color online) Variation of $I(s;x,y)$, $I(s;x)$ and $I(s;y)$ as a function of mean population of S, $\langle s\rangle$ for AA (a), AR (b), RA (c), and RR(d) motifs. Left and right column is for $n = 1$ and $n = 2$, respectively. The parameters are set at $\mu_s=0.1$ s$^{-1}$, $\mu_x=1.0$ s$^{-1}$, $\mu_y=10.0$ s$^{-1}$, $K_1=50$ (molecules/V), and $K_2=100$ (molecules/V). We use the mean copy number of X and Y at $\langle x\rangle=100$ (molecules/V), and $\langle y\rangle=100$ (molecules/V), respectively. The values of $k_s$, $k_x$, and $k_y$ are set according to the relations obtained by considering Langevin Eqs.~(\ref{eq1}-\ref{eq3}) at steady state. The lines are obtained from analytical calculations, while the symbols are due to numerical simulation using Gillespie algorithm \cite{Gillespie1976,Gillespie1977}.
}
\label{fig2}
\end{figure}


\section{Results and Discussion}

In the present article, we present a comparative study among the four TSC motifs with different architecture. Our goal here is to find out how architecture dependent information transmission is correlated with the fidelity of the networks. To this end, we tune the average input signal population $\langle s\rangle$ within the physiological range to study the proposed metrics of TSC motifs. To do so, we vary $k_s$, the synthesis rate constant of S, according to the relation $\langle s\rangle = k_s/\mu_s$, keeping the degradation rate constant $\mu_s$ fixed at a particular value. The mentioned relation can be obtained from Eq.~(\ref{eq1}) at the steady state.

\begin{figure}[!t]
\includegraphics[width=1.0\columnwidth,angle=0]{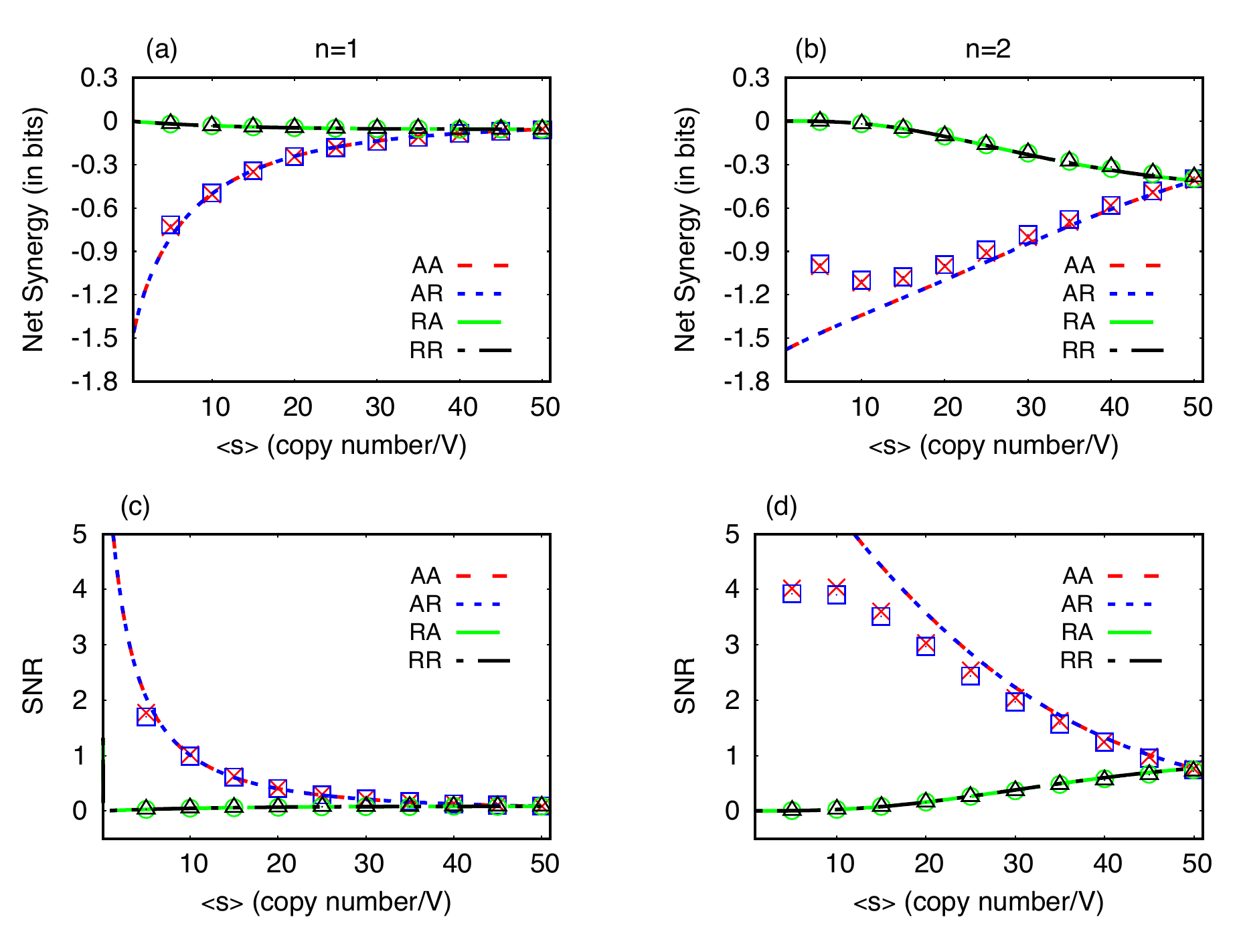}
\caption{(color online) Net synergy and signal-to-noise ratio as a function of mean population of S, $\langle s\rangle$. The left and right panels are for $n = 1$ and $n = 2$, respectively.The parameters used are $\mu_s=0.1$ s$^{-1}$, $\mu_x=1.0$ s$^{-1}$, $\mu_y=10.0$ s$^{-1}$, $K_1=50$ (molecules/V), and $K_2=100$ (molecules/V). We use the mean copy number of X and Y at $\langle x\rangle=100$ (molecules/V), and $\langle y\rangle=100$ (molecules/V), respectively. The values of $k_s$, $k_x$, and $k_y$ are set according to the relations obtained by considering Langevin Eqs.~(\ref{eq1}-\ref{eq3}) at steady state. The lines are obtained from analytical calculations, while the symbols are due to numerical simulation using Gillespie algorithm \cite{Gillespie1976,Gillespie1977}.
}
\label{fig3}
\end{figure}

We first quantitate the two variable and three variable MIs - $I(s;x)$, $I(s;y)$, and $I(s;x,y)$ using Shannon's definition for $n = 1$ and $n = 2$. These three MIs are plotted against the mean copy number of S, $\langle s\rangle$ in Fig.~\ref{fig2} for all four motifs. We observe a decreasing trend of MIs for AA and AR motifs and an increasing trend for RA and RR motifs. Validity of DPI and Markov chain property, i.e., $I(s;x,y) \approx I(s;x)$ and $I(s;x) > I(s;y)$ of the TSC motifs can be verified from Fig.~\ref{fig2}. We note that the profiles for AA and AR are the same irrespective of the architectural difference between the two motifs. RA and RR motifs also show similarity in their MI profiles. This observation suggests that the information transduction along the cascades depends only on the regulation S~$\rightarrow$~X but not on X~$\rightarrow$~Y. We also quantify net synergy and SNR for all motifs as a function of $\langle s\rangle$ (Fig.~\ref{fig3}).

It is important to mention that for $n = 2$ we observe disagreement between theoretical profiles and simulation results at low $\langle s \rangle$ (see Fig.~\ref{fig2}e-h and Fig.~\ref{fig3}b,d). As mentioned in the previous section $n > 1$ incorporates higher order nonlinearity in the system which also becomes prominent at low $\langle s \rangle$. The higher value of the Hill coefficient ($n>1$) takes into account the cooperative binding of TF to its binding site. This in turn brings in higher order nonlinearity in the dynamics due to which the network species experiences greater fluctuations. These fluctuations are more prominent when the mean population of signal $\langle s\rangle$ is low. This can be understood from Fig.~\ref{fig3}b,d which shows that at low $\langle s\rangle$ simulation results deviate from the theoretical profiles. For AA and AR motifs, such deviation is large. The deviations may even get larger in the case of $n \ge 2$ due to the breakdown of LNA.


Figure~\ref{fig3}a,b shows the values of net synergy are on the negative region, indicating the fact that this metric quantifies the net redundant information expressed by the output Y. This is further confirmed from Eq.~(\ref{eq7}), which shows that the net synergy value is equal to the negative magnitude of $I(s;y)$. The net synergy profiles for AA and AR fall on each other, and so are RA and RR. This behavior is similar to the MI profiles shown in Fig.~\ref{fig2}. The profiles for AA and AR are increasing while RA and RR are decreasing, and both become equal at high $\langle s\rangle$. To account for all these behavior, we focus on the analytical expression of SNR as the expression of net synergy has a logarithmic function (see Eq.~(\ref{eq7})). Using the expressions of various moments listed in Table~II, we have expressions of SNR for four different motifs (for $n = 1$)
\begin{eqnarray}
\label{eq8}
{\rm SNR_{AA}} & = &
\frac{B_1^2}{\alpha_1 + (\langle y\rangle+F_1) \langle s\rangle (K_1+\langle s\rangle)^2}, \\
\label{eq9}
{\rm SNR_{AR}} & = &
\frac{B_2^2}{\alpha_2 + (\langle y\rangle+F_2) \langle s\rangle (K_1+\langle s\rangle)^2}, \\
\label{eq10}
{\rm SNR_{RA}} & = &
\frac{B_3^2 \langle s\rangle}
{\alpha_3 \langle s\rangle + (\langle y\rangle+F_3) (K_1+\langle s\rangle)^2}, \\
\label{eq11}
{\rm SNR_{RR}} & = &
\frac{B_4^2 \langle s\rangle}
{\alpha_4 \langle s\rangle + (\langle y\rangle+F_4) (K_1+\langle s\rangle)^2}.
\end{eqnarray}

\noindent We refer to Appendix for detailed derivation of the above expressions of SNR along with the explicit forms of $\alpha_i$-s, $B_i$-s and $F_i$-s ($i = 1, 2, 3, 4$). These expressions have been plotted in Fig.~\ref{fig3}c,d, which shows similar collating phenomena of AA \& AR and RA \& RR, as observed for net synergy. The profiles shown in Fig.~\ref{fig3}c,d are obtained in the limit $K_1 \geqslant \langle s\rangle$. We further consider $K_1 + \langle s \rangle \approx K_1$ and approximate Eqs.~(\ref{eq8}-\ref{eq11}) leading to
\begin{eqnarray}
\label{eq12}
{\rm SNR_{AA}} & \approx &
\frac{B_1^2}{\alpha_1 + K_1^2 (\langle y\rangle + F_1) \langle s\rangle}, \\
\label{eq13}
{\rm SNR_{AR}} & \approx &
\frac{B_2^2}{\alpha_2 + K_1^2 (\langle y\rangle + F_2) \langle s\rangle}, \\
\label{eq14}
{\rm SNR_{RA}} & \approx &
\frac{B_3^2 \langle s\rangle}
{\alpha_3 \langle s\rangle + K_1^2 (\langle y\rangle + F_3)}, \\
\label{eq15}
{\rm SNR_{RR}} & \approx &
\frac{B_4^2 \langle s\rangle}
{\alpha_4 \langle s\rangle + K_1^2 (\langle y\rangle + F_4)}.
\end{eqnarray}

\noindent
From Appendix, it is clear that $B_1 = B_2$, $\alpha_1 = \alpha_2$, and $F_1 = F_2$, since in generating the results we have used, $\langle x\rangle = K_2$ (see all figure captions). This makes ${\rm SNR_{AA}}={\rm SNR_{AR}}$. Similarly, it can also be established that ${\rm SNR_{RA}}={\rm SNR_{RR}}$. This explains the similarity between AA and AR and between RA and RR. As mentioned earlier, the information transmission depends only on the regulation of S to X and not on the regulation of X to Y. This fact can also be understood from Eqs.~(\ref{eq12}-\ref{eq15}), which contain only $K_1$ as a signature of S to X regulation. However, the effect of X to Y regulation is encapsulated in the parameter $K_2$, which we surpass by making it equal with $\langle x\rangle$.


\begin{figure}[!t]
\includegraphics[width=1.0\columnwidth,angle=0]{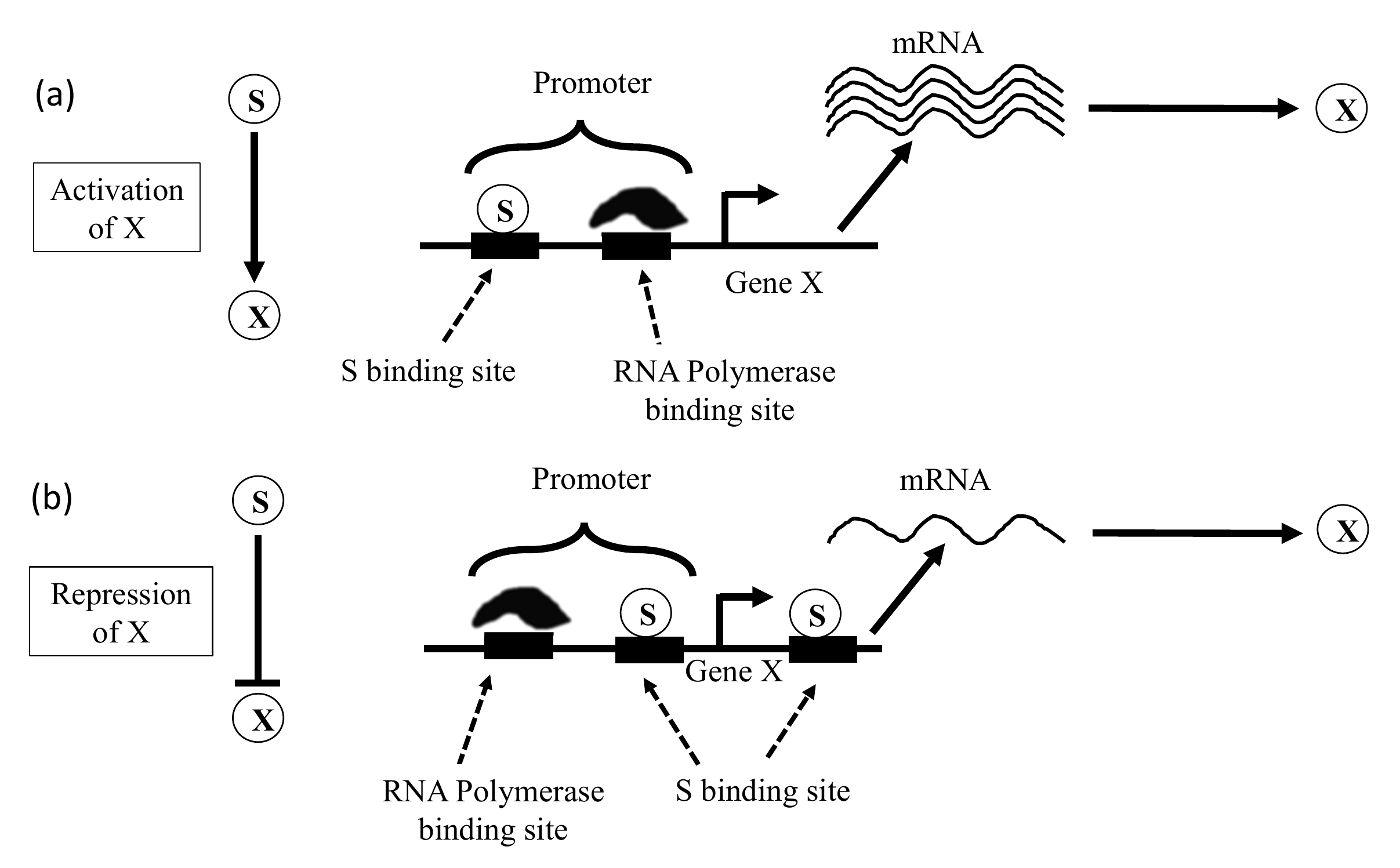}
\caption{Schematic diagram of (a) activation and (b) repression of X by the signaling molecule S.
}
\label{fig4}
\end{figure}

It is clear from Eqs.~(\ref{eq12}-\ref{eq13}) that, SNR for AA and AR is a decreasing function of $\langle s\rangle$, while Eqs.~(\ref{eq14}-\ref{eq15}) indicate that SNR for RA and RR is, indeed, an increasing function of $\langle s\rangle$. Such dependency on $\langle s \rangle$ is consistent with the simulation result shown in Fig.~\ref{fig3}c,d. These variations are significant only at low $\langle s\rangle$. At high $\langle s\rangle$, SNR of all motifs become equal. Since net synergy is the negative logarithm of SNR, the profiles of SNR are replicated in the profiles of net synergy with an opposite trend due to the negative sign.

In an earlier study, it has been shown that the intermediate component X in a TSC motif possesses some hindrance ability. As a result, information flow from S to Y gets reduced as the signal strength increases \cite{Nandi2019}. Such information restriction is prominent in the case of AA and AR motifs. In both the motifs, the signal S activates the intermediate X, whose restriction ability increases as a function of signal strength. Hence the information content $I(s;y)$ decreases (see Fig.~\ref{fig2}) and consequently net synergy increases (see Fig.~\ref{fig3}a,b) as a function of $\langle s\rangle$. This enhanced restriction character also causes a reduction in signalling fidelity of AA and AR motifs (see Fig.~\ref{fig3}c,d). On the other hand, in the case of RA and RR motifs, S represses the synthesis of X. Hence increasing signal strength here may decrease the restriction ability of X, which indeed results in elevated information transmission from S to Y, i.e., $I(s;y)$ increases with $\langle s\rangle$ (see Fig.~\ref{fig2}). Reduction in restriction ability of X, again, in turn, decreases the net synergy and increases the fidelity of RA and RR motifs (see Fig.~\ref{fig3}).

The above analysis indicates that the information transmission capacity along the four TSC motifs not only depends on the signal strength but also depends on the architecture of the motifs. In particular, the activation and repression phenomena also play an essential role. However, at a high signal population, such interplay of activation and repression does not remain valid, and all the motifs behave similarly. We have mentioned earlier that the results are presented in the limit $K_1 \ge \langle s\rangle$, which indicates that at $\langle s\rangle=50$ half-maximum point is reached (since for each plot we set $K_1 = 50$). We set the parameter $K_1$ in such a way that the extreme right points in each plot correspond to the half-maximum; whatever be the value of $\langle s\rangle$, up to which the proposed metrics are calculated. The convergence of net synergy and SNR continues even for $\langle s\rangle > 50$ by setting $K_1= {\rm maximum~ of}~ \langle s\rangle$. We note that if one crosses the half-maximum (by setting $K_1=50$ and by tuning $\langle s\rangle$ beyond 50), such convergence will break down and a crossover will occur.

Here it is imperative to understand the biological relevance of increasing $\langle s\rangle$ on the proposed metrics. We note here that we focus only on the activation-repression processes for S to X regulation. This is, perhaps, because the information transduction along the cascades depends only on S to X regulation. The signal S activates gene $X$ by binding at the promoter region upstream of RNA polymerase (RNAP) binding site (see Fig.~\ref{fig4}a). Consequently, the TF (or the signal) S together with RNAP stimulates the transcription processes so that protein X gets produced. On the other hand, S binds to the site downstream of RNAP binding site to inhibit the process of transcription (see Fig.~\ref{fig4}b). As a result, S represses the production of protein X.

When the signal population is high, S has a greater probability to get attached to their binding sites to either activate or inhibit the process of transcription. Such increased probability may lead to the incorporation of less fluctuations into the system of AA and AR. In consequence, information content $I(s;y)$ becomes less since information transmission along a network deciphers the propagation of the fluctuations. We note here that these fluctuations capture the knowledge of the concentration changes at the signal level \cite{Nandi2018}. Hence, net synergy increases for both AA and AR. On the other hand, increased binding probability of S leads to suppression of transcription rates in RA and RR. In other words, a reduced transcription rate may imply more randomness in the output copy number. Therefore, $I(s;y)$ increases, and net synergy decreases with $\langle s\rangle$. All these ups and downs in information transmission, in turn, affect the fidelity of the networks, as discussed previously.

\section{Conclusion}

In the present work, we studied information transmission in four different architecture of TSC motifs. Using the formalism of LNA, we have calculated various two variable and three variable MIs, net synergy, and SNR in terms of variance and covariance terms associated with the networks variables. We approximate these variables to follow Gaussian distribution. We also validated the proposed metrics by numerical simulation using Gillespie algorithm.

Quantification of MIs shows validation of DPI and Markov chain properties for four linear architectures having different interaction topology. We further validate our finding of redundancy in information transmission using the formalism of PID. Signal fidelity is quantified using the measure SNR. The variations in SNR can well explain the behavior of net synergy, as the latter is a logarithmic function of the former. From the approximate expressions of SNR, the clustering phenomena are understood.

The decreasing (of AA and AR) and increasing (of RA and RR) nature of different metrics with signal population are consistent with the approximate analytical expressions of SNR. We also postulate that the information transmission in these motifs depends only on the regulation of S to X. In other words, when the regulation is activating in nature, information transmission capacity goes down with increasing signal population. On the other hand, for repressive regulation, information transmission capacity increases with the signal population. Our analysis thus suggests that in a generic TCS motif, X to Y regulation has an insignificant role in information transmission. It has also been observed that the profiles of net synergy and SNR for AA and AR, and RA and RR differ significantly at low signal population. However, when the signal population is high, all the architectures have almost equal magnitudes of net synergy as well as SNR. At a high signal population, the interplay between activation and repression is no more a determining factor in the information dynamics of TSC motifs.

The four TSC motifs we considered in this work have simple architecture and can be constructed using genetic engineering and recombinant technology. To the best of our knowledge, only the RR motif has been constructed synthetically to study response delays in a transcription network of \textit{E. coli} \cite{Rosenfeld2003}. Using the similar principle of synthetic biology, one can construct the rest of the motifs (AA, AR, and RA) and experimentally verify signal fidelity in linear architectures with different interaction topologies. Our results demonstrate that a greater amount of redundant information in comparison with its synergistic counterpart ensures a significant proportion of the SNR or signal fidelity in the TSC irrespective of its topological specifics as well as the signaling strength

 
\begin{acknowledgements}
T.S.R. and M.N. thank CSIR, India [09/015(0495)/2016-EMR-I] and UGC, India [22/06/2014(i)EU-V], respectively, for research fellowships. 
\end{acknowledgements}


\section*{Appendix: Calculation of SNR}

For AA we have from Table~II
\begin{eqnarray*}
\Sigma(s,x) &=& \frac{k_x K_1 \langle s\rangle}{(K_1+\langle s\rangle)^2 (\mu_s+\mu_x)}, \\
&=& \frac{\mu_x \langle x\rangle (K_1+\langle s\rangle)}{\langle s\rangle} 
\times
\frac{K_1 \langle s\rangle}{(K_1+\langle s\rangle)^2 (\mu_s+\mu_x)}, \\
&=& \frac{K_1 \mu_x \langle x\rangle}{\mu_s+\mu_x}
\times
\frac{1}{K_1+\langle s\rangle}, \\
&=& \frac{A_1}{K_1+\langle s\rangle},
\end{eqnarray*}

\noindent where we have used $k_x=\mu_x \langle x\rangle (K_1+\langle s\rangle)/\langle s\rangle$ and $A_1=K_1 \mu_x \langle x\rangle/(\mu_s+\mu_x)$ for. Again using the expression of $\Sigma(s,x)$ and $k_y=\mu_y \langle y\rangle (K_2+\langle x\rangle)/\langle x\rangle$, we rewrite the expression of $\Sigma(s,y)$,
\begin{eqnarray*}
\Sigma(s,y) &=& \frac{k_y K_2 \Sigma(s,x)}{(K_2+\langle x\rangle)^2 (\mu_s+\mu_y)}, \\
&=& \frac{\mu_y \langle y\rangle (K_2+\langle x\rangle)}{\langle x\rangle} 
\times
\frac{K_2}{(K_2+\langle x\rangle)^2 (\mu_s+\mu_y)} \\
&& \times
\frac{A_1}{K_1+\langle s\rangle}, \\
&=& \frac{K_1 K_2 \mu_x \mu_y \langle y\rangle}{(K_2+\langle x\rangle) (\mu_s+\mu_x) (\mu_s+\mu_y)}
\times
\frac{1}{K_1+\langle s\rangle}, \\
&=& \frac{B_1}{K_1+\langle s\rangle},
\end{eqnarray*}

\noindent where $B_1=K_1 K_2 \mu_x \mu_y \langle y\rangle/((K_2+\langle x\rangle) (\mu_s+\mu_x) (\mu_s+\mu_y))$. It is important to note here that, the expressions of $k_x$ and $k_y$ are obtained by considering the Langevin equations corresponding to the motif at steady state. The expressions of other second moments are obtained by similar fashion and are given by,
\begin{eqnarray*}
\Sigma(x) &=& \langle x\rangle + \frac{C_1}{\langle s\rangle (K_1+\langle s\rangle)^2}, \\
\Sigma(x,y) &=& D_1 + \frac{E_1}{\langle s\rangle (K_1+\langle s\rangle)^2}, \\
\Sigma(y) &=& (\langle y\rangle + F_1) + \frac{G_1}{\langle s\rangle (K_1+\langle s\rangle)^2},
\end{eqnarray*}

\noindent with
\begin{eqnarray*}
C_1 &=& \frac{K_1^2 \mu_x \langle x\rangle^2}{\mu_s+\mu_x}, \\
D_1 &=& \frac{K_2 \mu_y \langle y\rangle}{(K_2+\langle x\rangle)(\mu_x+\mu_y)}, \\ 
E_1 &=& \frac{K_1^2 K_2 \mu_x \mu_y \langle x\rangle \langle y\rangle (\mu_s+\mu_x+\mu_y)}
{(K_2+\langle x\rangle)(\mu_s+\mu_x)(\mu_s+\mu_y)(\mu_x+\mu_y)}, \\
F_1 &=& \frac{K_2^2 \mu_y \langle y\rangle^2}{\langle x\rangle (K_2+\langle x\rangle)^2 (\mu_x+\mu_y)}, \\
G_1 &=& \frac{K_1^2 K_2^2 \mu_x \mu_y \langle y\rangle^2 (\mu_s+\mu_x+\mu_y)}
{(K_2+\langle x\rangle)^2(\mu_s+\mu_x)(\mu_s+\mu_y)(\mu_x+\mu_y)}.
\end{eqnarray*}

\noindent Now utilizing the expressions of different moments one can express SNR associated with the motif AA in terms of model parameters and $\langle s \rangle$ as
\begin{equation*}
{\rm SNR_{AA}} =
\frac{B_1^2}{\alpha_1 + (\langle y\rangle+F_1) \langle s\rangle (K_1+\langle s\rangle)^2},
\end{equation*}

\noindent with $\alpha_1=G_1 - B_1^2$. At low population of the signal S, we approximate that $K_1 \gg \langle s\rangle$ and hence $K_1+\langle s\rangle \approx K_1$. This makes the SNR to be a hyperbolic function of $\langle s\rangle$,
\begin{equation*}
{\rm SNR_{AA}} \approx \frac{B_1^2}{\alpha_1 + K_1^2 (\langle y\rangle+F_1) \langle s\rangle}.
\end{equation*}

For AR we evaluate the expressions of different moments as a function of $\langle s\rangle$ following the same strategy
\begin{eqnarray*}
\Sigma(s,x) &=& \frac{A_2}{K_1+\langle s\rangle}, \\
\Sigma(s,y) &=& -\frac{B_2}{K_1+\langle s\rangle}, \\
\Sigma(x) &=& \langle x\rangle + \frac{C_2}{\langle s\rangle (K_1+\langle s\rangle)^2}, \\
\Sigma(x,y) &=& -D_2 - \frac{E_2}{\langle s\rangle (K_1+\langle s\rangle)^2}, \\
\Sigma(y) &=& (\langle y\rangle + F_2) + \frac{G_2}{\langle s\rangle (K_1+\langle s\rangle)^2},
\end{eqnarray*}

\noindent where, $A_2=A_1$, $B_2=(\langle x\rangle/K_2)B_1$, $C_2=C_1$, $D_2=(\langle x\rangle/K_2)D_1$, $E_2=(\langle x\rangle/K_2)E_1$, $F_2=(\langle x\rangle/K_2)^2F_1$ and $G_2=(\langle x\rangle/K_2)^2G_1$. The expression of SNR becomes
\begin{equation*}
{\rm SNR_{AR}} = \frac{B_2^2}{\alpha_2 + (\langle y\rangle+F_2) \langle s\rangle (K_1+\langle s\rangle)^2}.
\end{equation*}

\noindent Here, $\alpha_2 = G_2 - B_2^2 = (\langle x\rangle/K_2)^2 \alpha_1$. At low $\langle s\rangle$ value, the SNR follows the same trend as was observed in case of AA motif.

For RA we write the expressions of second moments as,
\begin{eqnarray*}
\Sigma(s,x) &=& -A_3 \frac{\langle s\rangle}{K_1+\langle s\rangle}, \\
\Sigma(s,y) &=& - B_3 \frac{\langle s\rangle}{K_1+\langle s\rangle}, \\
\Sigma(x) &=& \langle x\rangle + C_3 \frac{\langle s\rangle}{(K_1+\langle s\rangle)^2}, \\
\Sigma(x,y) &=& D_3 + E_3 \frac{\langle s\rangle}{(K_1+\langle s\rangle)^2}, \\
\Sigma(y) &=& (\langle y\rangle + F_3) + G_3 \frac{\langle s\rangle}{(K_1+\langle s\rangle)^2},
\end{eqnarray*}

\noindent where,
\begin{eqnarray*}
A_3 &=& \frac{\mu_x \langle x\rangle}{\mu_s+\mu_x}, \\
B_3 &=& \frac{K_2 \mu_x \mu_y \langle y\rangle}{(K_2+\langle x\rangle) (\mu_s+\mu_x) (\mu_s+\mu_y)}, \\
C_3 &=& \frac{\mu_x \langle x\rangle^2}{\mu_s+\mu_x}, \\
D_3 &=& \frac{K_2 \mu_y \langle y\rangle}{(K_2+\langle x\rangle) (\mu_x+\mu_y)}, \\
E_3 &=& \frac{K_2 \mu_x \mu_y \langle x\rangle \langle y\rangle (\mu_s+\mu_x+\mu_y)}
{(K_2+\langle x\rangle) (\mu_s+\mu_x)(\mu_s+\mu_y)(\mu_x+\mu_y)}, \\
F_3 &=& \frac{K_2^2 \mu_y \langle y\rangle^2}{\langle x\rangle (K_2+\langle x\rangle)^2 (\mu_x+\mu_y)}, \\
G_3 &=& \frac{K_2^2 \mu_x \mu_y \langle y\rangle^2 (\mu_s+\mu_x+\mu_y)}
{(K_2+\langle x\rangle)^2 (\mu_s+\mu_x)(\mu_s+\mu_y)(\mu_x+\mu_y)}.
\end{eqnarray*}

\noindent Now, the expression of SNR is,
\begin{equation*}
{\rm SNR_{RA}} = \frac{B_3^2 \langle s\rangle}
{\alpha_3 \langle s\rangle + (\langle y\rangle+F_3) (K_1+\langle s\rangle)^2},
\end{equation*}

\noindent with, $\alpha_3 = G_3 - B_3^2$. At low $\langle s\rangle$, SNR will become,
\begin{equation*}
{\rm SNR_{RA}} \approx \frac{B_3^2 \langle s\rangle}
{\alpha_3 \langle s\rangle + K_1^2 (\langle y\rangle+F_3)},
\end{equation*}

\noindent which is an increasing function of $\langle s\rangle$.

On a similar note, for RR the expressions of second moments are,
\begin{eqnarray*}
\Sigma(s,x) &=& -A_4 \frac{\langle s\rangle}{K_1+\langle s\rangle}, \\
\Sigma(s,y) &=& B_4 \frac{\langle s\rangle}{K_1+\langle s\rangle}, \\
\Sigma(x) &=& \langle x\rangle + C_4 \frac{\langle s\rangle}{(K_1+\langle s\rangle)^2}, \\
\Sigma(x,y) &=& -D_4 - E_4 \frac{\langle s\rangle}{(K_1+\langle s\rangle)^2}, \\
\Sigma(y) &=& (\langle y\rangle + F_4) + G_4 \frac{\langle s\rangle}{(K_1+\langle s\rangle)^2},
\end{eqnarray*}

\noindent where $A_4=A_3$, $B_4=(\langle x\rangle/K_2)B_3$, $C_4=C_3$, $D_4=(\langle x\rangle/K_2)D_3$, $E_4=(\langle x\rangle/K_2)E_3$, $F_4=(\langle x\rangle/K_2)^2 F_3$ and $G_4=(\langle x\rangle/K_2)^2 G_3$. The expression of SNR is,
\begin{equation*}
{\rm SNR_{RR}} = \frac{B_4^2 \langle s\rangle}
{\alpha_4 \langle s\rangle + (\langle y\rangle+F_4) (K_1+\langle s\rangle)^2},
\end{equation*}

\noindent with $\alpha_4 = G_4 - B_4^2 = (\langle x\rangle/K_2)^2 \alpha_3$. At low $\langle s\rangle$, SNR shows similar trend as a function of $\langle s\rangle$ as was observed for RA motif.


%

\end{document}